\documentclass[12pt, reprint, aps, prb]{revtex4-2} 

\usepackage{graphicx, amsmath, amssymb}
\setcitestyle{super}
\begin{document}

\title{Letter to the Editor:  Symmetry and Voltmeters}

\maketitle 

Consider a loop made of two equal resistors that is concentric with an ideal, infinite solenoid.  
Let a changing current in the solenoid induce an electromotive force in the loop.  
Attach two identical voltmeters in exactly the same manner to the loop in parallel with the resistors, but place them on different sides of the solenoid.  
What do the voltmeters read? 
This experiment and its variations make engaging demonstrations, and are the subject of many articles in this journal\cite{nicholson:2005, young:1988, peters:1984, phillips:1983, romer:1982, reif:1981, hart:1981, klein:1981, moorcroft:1970, moorcroft:1969} and elsewhere.\cite{zangara:1995, zik:1991, phillips:1963, buchta:1963}  

At first, the fact that the two voltmeters give different readings is startling.\cite{jackson:2011}  
A quick way to understand this result is to consider symmetry under rotations and reflections.  
Notice that a rotation about the solenoid's axis does not change the physical system (see Fig.~1).
However, a 180$^\circ$ rotation effectively swaps the positions and reverses the polarities of the voltmeters in Fig.~1.   
Thus, the readings must satisfy $V_1 = - V_2$.  
Similarly, a reflection (or 180$^\circ$ rotation) about the line $AB$ swaps the voltmeters' positions and reverses the solenoid's current (instead of the voltmeters' polarities) in Fig.~1, giving the same result.

In contrast, if a battery bridges the loop in parallel with the resistors and there is no solenoid, 
symmetry shows that the readings must be equal. 
\\ 

\begin{figure}[htbp]
\begin{center}
\includegraphics[width=0.4\textwidth]{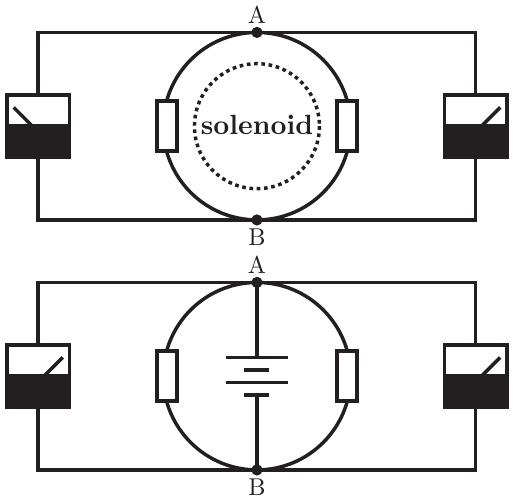}
\caption{Schematic diagrams of the circuits.}
\label{default}
\end{center}
\end{figure}

\begin{flushright}
Bart H. McGuyer

Graduate Student

Department of Physics

Princeton University

\end{flushright}

\end{document}